\documentclass{article}
\usepackage{epsf,epsfig}
\title{{\bf Some doubts on the validity of the foreground Galactic 
contribution subtraction from microwave anisotropies}}

\author{Mart\'\i n L\'opez-Corredoira\\
Instituto de Astrof\'\i sica de Canarias\\ C/.V\'\i a L\'actea, 
s/n\\ E-38200 La Laguna (S/C de Tenerife), Spain\\ 
E-mail: martinlc@iac.es}

\begin{document}
\maketitle

\vspace{60mm}

24 pages, 1 figure

\eject

ABSTRACT:

The Galactic foreground contamination in CMBR anisotropies, especially
from the dust component, is not easily separable from the
cosmological or extragalactic component.
In this paper, some doubts will be raised concerning the validity of
the methods used to date to remove Galactic dust emission
in order to show that none of them achieves its goal.

First, I review the recent bibliography on the topic and discuss 
critically the methods of foreground subtraction: 
the cross-correlation with templates, analysis assuming the
spectral shape of the Galactic components, the ``maximum entropy
method'', ``internal linear combination'', 
and ``wavelet-based high resolution fitting of internal templates''. 
Second, I analyse the galactic latitude dependence 
from WMAP data. The frequency dependence
is discussed with the data in the available literature. The result
is that all methods of subtracting the Galactic contamination are inaccurate.
The galactic latitude dependence analysis or the frequency dependence 
of the anisotropies in the range 50--250 GHz put a constraint on
the maximum Galactic contribution in the power spectrum 
to be less than a $\sim$10\% (68\% C. L.) for a $\sim$1 degree scale,
and possibly higher for larger scales.

The origin of most of the signal in the CMBR anisotropies is not
Galactic. In any case, the subtraction of the Galaxy is not accurate enough to 
allow a ``precision Cosmology''; other sources of contamination
(extragalactic, solar system) are also present.

Keywords: cosmic microwave background -- ISM: Clouds -- 
   dust, extinction -- ISM: structure
   

\eject

\section{Introduction}

The cosmic microwave background radiation (CMBR) has been interpreted as the relict radiation
of an early stage of the Universe. Its black-body spectrum (Mather et al. 1994) 
of 2.75 K reveals a very small dependence on sky
position. Measurements of the anisotropies,
carried out by several teams of researchers over the last two decades,
have been claimed to provide information on the structural formation of 
the Universe, inflation in its early stages, quantum gravity, topological 
defects (strings, etc.), dark matter type and abundance, the determination of 
cosmological parameters ($H_0$, $\Omega _m$, $\Omega _\Lambda $, etc.), 
the geometry and dynamics of the Universe, the thermal history of 
the Universe at the recombination epoch, etc. (e.g., Bennett et al. 2003a). 
There are even some researchers who claim
that the golden era of cosmology, the age of the precision 
cosmology, has arrived with 
the new experiments to measure these anisotropies and other optimistic
interpretations of observational cosmology.
However, all that glitters is not gold and
what is claimed as a source of cosmological information is often something else.
One should be very careful to ensure that there are no contaminants
along the light path of this microwave radiation, 
Galactic contamination being one important factor. 
If one wants to do ``precision cosmology'', then even greater care must be taken. 
Many authors (see the brief review in \S \ref{.status}) have
studied the different components of the Galactic microwave foreground 
radiation, but in my view these efforts are still insufficient to separate 
the components appropriately. The foreground emission is small;
there is no {\it ``galactic foreground contamination which is 
1,000 times more intense than the desired signal´´} (as claimed by 
Robitaille 2007) in off-plane regions; this is completely wrong,
but neither  is the foreground contamination entirely negligible. 
At least, some doubts on the validity of the foreground Galactic 
subtraction from microwave anisotropies can be expressed, and this is
precisely the topic of this paper, together with some analysis
to constrain Galactic emission.
In \S \ref{.status}, I review the recent bibliography on the topic and discuss 
critically the methods of foreground subtraction.
I analyse the galactic latitude dependence (\S \ref{.galdep}) from WMAP
first year data. 
The frequency dependence
is discussed (\S \ref{.freqdep}) with the data in the available literature.
Other possible sources of contamination are also discussed in
\S \ref{.others}.

\section{Some comments on the present status of CMBR foreground analysis}
\label{.status}

The consideration of the Galactic foregrounds has been present for a long time now.
However, an analysis of the papers analysing the problem reveals
that it was sometimes underestimated,  and that the problem of 
Galaxy subtraction has become more complex in recent years.

For instance, Guti\'errez de la Cruz et al. (1995), Davies et al. (1996) 
thought that Tenerife data at 15 GHz (see also Guti\'errez et al.
2000) were likely to be dominated by 
cosmological fluctuations, and that 33 GHz data were almost unaffected 
by the Galaxy ($\sim 4\mu $ K in scales of 5--15 degrees). More recent analyses
(WMAP, Bennett et al. 2003b, Fig. 10), however, have claimed that for comparable
angular scales the Galaxy is dominant in the anisotropies at 22.8 GHz 
(and the Galactic contamination at 15 GHz is not much smaller than at 22.8 GHz), 
and at 33 GHz the Galactic anisotropies are of the same order as
those from the CMBR. Also, at small scales, less than a degree, the foreground
Galactic emission dominates (Leitch et al. 2000).
Halverson et al. (2002) and 
Fern\'andez-Cerezo et al. (2006) claim that the cosmological signal
is still dominant at $|b|>40^\circ $ around 15 GHz and 31 GHz
respectively for intermediate and small
angular scales (around a degree and down to $\approx 12$ 
arcminutes respectively), 
but their analysis is based on correlations with templates, which,
as we will see below, might not be appropriate.
An explanation for this is that it is mainly due to the
existence of a new kind of emission correlated with the dust that
was not discovered previously;  the 
synchrotron and or free-free emission might also have been underestimated. Positive correlations between the 
microwave anisotropies, including the region around 15 GHz, 
and far-infrared maps, which trace Galactic dust, 
were found (Kogut et al. 1996a; Leitch et al. 1997, 2000; 
de Oliveira-Costa et al. 1997, 1998; Finkbeiner et al. 1999, 2004;
Fern\'andez-Cerezo et al. 2006).
Casassus et al. (2004) report that in the Helix region 
the emission at 31 GHz and 100$\mu $m are well correlated; they 
estimate that the 100$\mu $m-correlated radio emission, 
presumably due to dust, accounts for at least 20\% of the 31 GHz emission 
in the Helix (so the total dust emission is higher than 20\% because there
is a non-correlated component too).
Watson et al. (2005) show that the anomalous emission at around 23 GHz
of the Perseus molecular cloud (Temp$\sim 1$ mK)
is an order of magnitude larger than the emission expected from synchrotron +
free--free + thermal dust.
There is also some correlation of 10, 15 GHz maps with H$_\alpha$ maps,
but very low, so the free--free emission is detected at levels far lower
than the dust correlation (de Oliveira-Costa et al. 1999, 2002).
The most likely current explanation for this emission correlated with
dust around 15--50 GHz is spinning dust grain emission (Draine \& Lazarian 
1998) and/or magnetic dipole emission from ferromagnetic grains 
(Draine \& Lazarian 1999); although some 
other authors (e.g., Mukherjee et al. 2001) do not agree.
Whatever it is, it is now clear that the 15--40 GHz range
 is dominated by the Galaxy.

A question might arise as to whether the contamination in the remaining
frequencies (40--200 GHz) is being correctly accounted for.
Typical calculations of dust contamination claim that it should not be 
predominant, and that its contribution can be subtracted accurately,
but how sure can we be of this statement? How accurate is the
subtraction of the dust foreground signal?

\subsection{Dust foreground removal with templates}

A first difficulty in the subtraction of the dust component is to
know exactly how much  emission there is in each line of sight.
First approximations  came with
extrapolations from the IRAS far-infrared and  DIRBE data 
(with the zodiacal light subtracted, as well the cosmic infrared background) 
used to model the dust thermal emission in microwaves.
Templates were taken from these infrared maps and 
extrapolated in amplitude by a common factor for all pixels.
The problem is that, as said by Finkbeiner et al. (1999), {\it `a template 
approach is often carelessly used to compare 
observations with expected contaminants, with the correlation amplitude 
indicating the level of contamination. (...) These templates ignore 
well-measured variation in dust temperature and variations in 
dust/gas ratio.'}
According to L\'opez-Corredoira (1999), the growing contrast of colder
clouds in the background of the diffuse interstellar medium will
produce much higher microwave anisotropies than the product of the
template extrapolation. 
Neither  is it a good strategy
to subtract a scaled IRAS template to remove the spinning dust
in multifrequency data since it produces large residual differences
(Leitch et al. 2000).

A better approximation to  this dust emission in the microwave region
came with the adoption of an extrapolation with colour corrections in each pixel.
This method assigns a different temperature to each pixel. This colour correction,
together with a $\nu ^2$ emission emissivity (Schlegel et al. 1998),
gave a much tighter agreement with FIRAS data. However, it
was still inconsistent with the FIRAS data below 800 GHz 
in amplitude (Finkbeiner et al. 1999): a 14\% error in the mean amplitude 
at 500 GHz. Indeed, no power-law
emissivity function fits the FIRAS data in the 200--2100 GHz region (Finkbeiner 
et al. 1999). Furthermore, laboratory measurements suggest that the
universality of $\nu ^2$ emissivity is an oversimplification, with
different species of grains having different emissivity laws 
(Finkbeiner et al. 1999). A better approximation is an extrapolation
with two components ($\nu ^{1.7}$, $\langle T\rangle $=9.5 K and $\nu ^{2.7}$,
$\langle T\rangle $=16 K); each pixel has an assignation of two temperatures 
(Finkbeiner et al. 1999) with correction factors that are a function of 
these temperatures. This still fails in the predictions of
FIRAS from the IRAS--DIRBE extrapolation by 15\% in zones dominated 
by atomic gas (Finkbeiner et al. 1999).

Finkbeiner et al.'s (1999) approach, although
much better than a direct extrapolation of the template, is insufficient.
The problem is difficult to solve because we are trying to do
an extrapolation by a factor $\sim 15-60$ in frequency (from 1250 GHz
[240 $\mu $m] or 3000 GHz [100 $\mu $m]) to 50--90 GHz). This is equivalent,
for instance, to the attempt to derive a map of stellar emission in
12$\mu $m as an extrapolation of the emission in optical B-filter.
Frankly, when I see IRAS-12$\mu $m map of point sources and
the Palomar plates in blue filters, I observe huge differences, 
and I do not know how we can extrapolate
the second map to obtain the first one. Each star has a different
colour, and stars which are very bright in blue may be very
faint at mid-infrared and vice verse. The same thing happens with the
diffuse + cloud emission: there are hot regions, cold regions, different kinds
of emitters (molecular gas, atomic gas) and we have to integrate all
this into each line of sight. The assumption that with only two temperatures
we can  extrapolate the average flux, and that a colour term can
correct the pixel-to-pixel differences is comparable to the assumption that 
a model with only two kinds of stars and the knowledge of $(B-V)$ for each star
we can extrapolate the star counts from optical to mid-infrared 
for the whole Galaxy.

The cold clouds, invisible or very faint in far-infrared surveys, are
potential elements to produce a very significant emission at 40--200 GHz
at the amplitude level of the observed anisotropies.
Bernard et al. (1999) have shown that some cirrus features with a high value of 
$I_{100\mu m}/I_{60\mu m}$ of 29.5 (the average is 3.2) are
cold dust regions ($T\approx 13$ K instead of the average of 17.5 K).
Colder clouds may also exist.
Lawrence (2001) claims that part of the SCUBA-850 $\mu m$ sources might be
local sources at 7 K. And who knows whether there are even colder clouds
that might emit significantly only in the range of microwaves?
Paladini et al. (2007)
show that the microwave emission in the outer Galactic plane ($R>8.9$ kpc)
is much higher than expected, suggesting the possible presence of
very cold dust.

It is also very common (e.g., Masi et al. 2001, Fern\'andez-Cerezo et al.
2006) to calculate the Galactic dust contribution
in some microwave data by just making a cross-correlation between
these data and some far-infrared map of the sky. This is simply
wrong and not even valid for ascertaining the order of magnitude of
such contamination. Since the templates in infrared cannot be used
as maps of the microwave Galactic emission, there will be non-correlated
emission coming from the dust too. That is, there will be many clouds
that emit substantially  in microwaves, and that are not detected
in far-infrared maps.

Therefore, any method of foreground subtraction which uses templates
will have serious credibility problems with regard to the goodness of the
subtraction. This applies not only to those methods that use templates directly with
coupling coefficients derived from cross-correlation
but also to MEM (maximum entropy method; Bennett et al. 2003b), which uses
in the initial stage templates for the dominant foreground components
and also establishes some a priori conditions of their spectral behaviour.
Moreover, any calculation of the limits of such contamination based on
cross-correlations will not be totally accurate.

\subsection{Dust foreground removal without templates}

If removal of foreground contamination is  applied without the use
of templates, we might need a knowledge of all of the physical components of
this foreground emission and their spectral behaviour. 

With ISO-90,180$\mu $m, a power law down to scale of 3$'$ is found in
the power spectrum (Herbstmeier et al. 1998), 
with similar shape as that derived from the 21 cm line for Galactic cirrus. 
At 410 GHz, the dust emission is also roughly a power law in off-plane
regions (Masi et al. 2001), although I see in the plot for $b=-17^\circ$
from  figure 2 of Masi et al. (2001) that there would be a better fit
if we admitted that there was a maximum in the dust emission power spectrum 
around $l=200$. In any case, even if
a power law is present in the far infrared or at 410 GHz, the extrapolation
to 50--90 GHz is not direct and, moreover,
the spectral index of this power law is highly variable.
The presence of different dust-emitting components causes 
the spectral indices of the foregrounds to vary with position.
Kiss et al. (2003) have measured spectral index changes 
from $-$5.1 to $-$2.1, depending
on the region, and also changes with wavelength; they
point out the existence of dust at different temperatures,
in particular of a cold, extended component.
Dupac et al. (2004) also show with the PRONAOS balloon-borne experiment that
the submillimetre/millimetre spectral index is found to vary between 
roughly 1 and 2.5 on small scales (3.5$'$ resolution).
Even small spectral index variations as small as $\Delta \alpha \sim 0.1$
can have a substantial impact on how channels should be combined and on the
attainable accuracy (Tegmark 1998), so the errors of the subtraction 
assuming certain power spectrum for the dust are serious.

Another technique that does not use templates is (Bennett et al. 2003b):
ILC (internal linear combination). It assumes nothing about
the particular frequency dependencies or morphologies of the foregrounds and
tries to minimize the variance in twelve different regions of the sky
with the combination of the five available WMAP frequencies.
There is degeneracy of solutions, an infinite number of maps can
be generated from the five basis sets, there is no way to test
whether the maximum likelihood solution is the correct one (Robitaille 2007),
and it is not enough to divide the sky into twelve regions.
Bennett et al. (2003b) themselves warn against its use for cosmological
analysis, and Eriksen et al. (2004b) show that
it is not effective to remove all residual foregrounds.
The ILC method performs quite badly, especially for dust (Eriksen et al. 2004b),
in part because of the variability of the spectral indices. Indeed, a
tremendous coefficient variability in the twelve sections was obtained
(Robitaille 2007). Therefore,
ILC maps are not clean enough to allow cosmological conclusions to be arrived at
(Eriksen et al. 2004b): {\it `[The] ILC map, which by eye looks almost 
free of foreground residuals, has been extensively used for scientific
purposes---despite the fact that there are strong (and difficult to
quantify) residual foregrounds present in the map. (...) the ILC map
is indeed highly contaminated by residual foregrounds, and in particular,
that the low-$l$ components, which have received the most attention so
far, are highly unstable under the ILC cleaning operation'} (Eriksen
et al. 2005). 

The WI-FIT method (`Wavelet based hIgh resolution Fitting of
Internal Templates', Hansen et al. 2006) does not require a priori
templates, but takes the information about the foregrounds
by taking differences of temperature maps at different frequencies. 
However, for the application in presently available
maps, it requires the assumption that the spectral indices are
constant in space, which, as said, is a very inaccurate approximation.
Their assumption that the Galactic emission in each pixel is
proportional to the difference in temperature maps 
($(T_i^\nu - T_i^{\nu '})\propto T_i^\nu $) is in general 
incorrect because $T_i^{\nu '}$ is not proportional to $T_i^\nu $, 
the temperature in each pixel being the superposition of many different emissions
with different temperatures. Again, we have here the same problem
as with the use of templates: the assumption
that there is only one temperature along each line of sight,
and that the intensity of this emission is describable
with a simple average fixed power law multiplying a black body emission
with an average temperature. Hansen et al. (2006) claim that
their method is good because they obtain similar results to
Bennett et al. (2003b), but this may be due to their similar
assumptions.

Foreground contamination residuals are found even for the best
available supposed clean CMB maps. 
Naselsky et al. (2006) found some correlation between $\Delta l=4n$ and $n$=1,2
spherical harmonic multipole domain, which is caused by a symmetric 
signal in the Galactic coordinate system.
de Oliveira-Costa \& Tegmark (2006)
find that the alignment of low-l multipoles
appears to be rather robust to Galactic cut and different
foreground contaminations.
It is also relevant that Bernui et al. (2007a) found a preferred direction in 
the fluctuations of the WMAP data while
there should be no preferential direction in the sky (isotropy); but the
significance of this detection is only at 95\% C.L. so it could be
just a chance detection, or  it could also indicate some kind of contamination.
The statistical anisotropy in different circles of the sky
is also found by Then (2006). Liu \& Zhang (2006) have also studied the
cross-correlation between WMAP and Egret $\gamma $-ray data and concluded
than an unknown source of radiation, most likely of galactic origin, is
implied by their analysis. Verschuur (2007) speculate that this
unknown radiation of galactic origin might take place in the surface
of Galactic HI structures moving through interstellar space and/or
interacting with one another. 

The spatial association found by Verschuur (2007) on scales of 1--2 degrees 
between interstellar neutral hydrogen,
integrated in maps over ranges of 10 km/s, and WMAP-ILC maps,
which should be clean of foreground contamination through the ILC methods,
is especially significant for the present discussion too. Several extended
areas of excess emission at high galactic latitudes ($b>30^\circ $)
are present in both maps. According to Verschuur (2007), these structures
have typical distances from the Sun of order 100 pc.

\subsection{Non-gaussianity}
\label{.gauss}

Another aspect to think about is the non-gaussianity of the anisotropies
distribution. Standard theories involving inflation generally predict a pattern of
gaussian noise; whereas non-standard theories based on 
symmetry breaking and the
generation of defects have more distinctive signatures (Coulson et al. 1994).
The fact is that a non-gaussian distribution was discovered
in the anisotropies against all predictions of current inflationary
cosmological models by many different authors with different methods
(Ferreira et al. 1998; Pando et al. 1998; Magueijo 2000a,b;
Chiang et al. 2003; Coles et al. 2004; 
Eriksen et al. 2004a; Park 2004; Vielva et al. 2004; Liu \& Zhang 2005;
Raeth et al. 2007; Bernui et al. 2007b) 
and yet the leading cosmologists still 
claim inflationary cosmology to be valid  and to be able to constrain 
cosmological parameters from the anisotropies. 
This might mean that the analyses of
non-gaussianity are wrong, or that they are not significant enough, or that the
inflation model is incorrect, or that some contamination is present
in the maps, which are supposed to be clear of any contamination.
Among all these possibilities, maybe more than one applies, and I suspect
that one of these is the incorrect subtraction of the Galactic contamination
although other factors may also be important. As a matter of fact,
Liu \& Zhang (2005) show with  certain tests that residual foreground 
contamination in WMAP-cleaned maps may contribute to this non-gaussian features 
significantly, and Tojeiro et al. (2006) have argued that the non-gaussianity
is associated with cold spots of unsubtracted foregrounds.
The lowest spherical harmonic modes in the map are significantly 
contaminated with foreground radiation (Chiang et al. 2007).
However, galactic contamination may be not the only reason:
using the WMAP data and 2MASS galaxy catalog, Cao et al. (2006) show that the 
non-gaussianity of the 2MASS galaxies is imprinted on WMAP maps too.
Nonetheless, Rubi\~no-Mart\'\i n et al. (2006) and McEwen et al. (2006) 
claim that only some few regions have such a non-gaussian anisotropies due to 
contamination, e.g. the Corona Borealis supercluster region, and that most  
regions of the sky are gaussian.


\section{Galactic latitude dependence}
\label{.galdep}

A simple way to test whether the Galactic anisotropies are
somewhat significant would be examining their variation with the position in the
sky, for example with galactic latitude dependence (L\'opez-Corredoira
1999). Higher Galactic anisotropies are expected toward the plane.
Moreover, the lines of sight in the 
southern Galactic hemisphere should give more Galactic anisotropies
than the northern one,
because the Sun is approximately 15 pc above the Galactic plane  
(Hammersley et al. 1995; Chyzy et al. 2005)
and the scale height of the cold dust component is not very high, although
unknown in principle. For dust observed in infrared, the scale-height is 40 pc 
(Unavane et al. 1998)\footnote{Note that, although Unavane et al. use
near-infrared data, the scale-height of 40 pc does not refer to the
old stellar population (the scale-height of the old stellar
population is 285 pc; L\'opez-Corredoira et al. 2002)
but to the dust distribution inferred through the extinction maps.}.
If the cold dust responsible for the microwave
fluctuation in the range 60--100 GHz had a scale-height of 40 pc as well,
the excess column density in the south over the north 
$\left (\frac{2-\exp{\frac{-15}{40}}}{\exp{\frac{-15}{40}}}-1\right )$ 
would be 91\%, for any galactic latitude.
If we assume a cosecant law for galactic latitude dependence,
this absolute value of the asymmetry would be $\approx 3$ times higher
at $|b|=20^\circ $ than at $|b|=90^\circ $.
Of course, this is a rough calculation because the scale-height might change
in microwaves with respect to the value inferred from the extinction in 
infrared, there is longitude dependence, flares, the accuracy of the cosecant 
law for low latitude regions is not good. Clouds sizes and other factors 
should be taken into account too.

Nonetheless, we are limited by the variations of the anisotropies 
produced by the own cosmological anisotropies, especially those at
large scale, which have a range of possible gradients
that could even mimic the Galactic gradients. 
Multipoles of low-l with $l\ge 2$ are part of the fluctuations,
and one might suspect that they are responsible for the gradients and
the difference north/south.
One way to check whether these variations with Galactic position exist
and whether they are within the expected cosmological variation or not is
to measure them with real data (WMAP) and compare with realizations.

\subsection{WMAP data analysis}
 
\begin{figure}
\begin{center}
\mbox{\epsfig{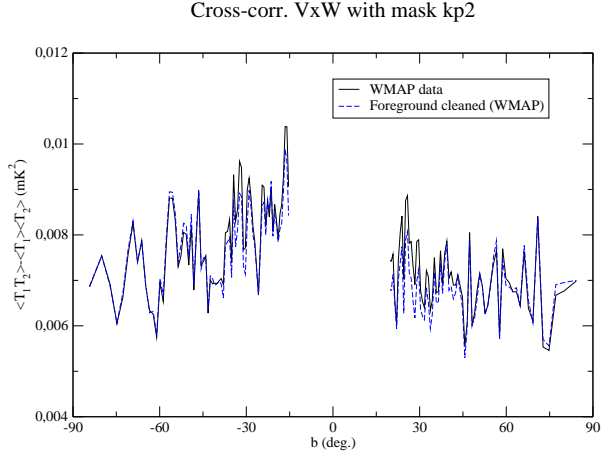}}
\end{center}
\caption{Variance as a function of galactic latitude in WMAP 
first-year-data (only points with less than 10\% of pixels in the
mask kp2). Each point in the plot represents the pixels
in the ring for all galactic longitudes 
and $\Delta (\sin b)=0.01$.}
\label{Fig:crossvarVW}
\end{figure}

Figure \ref{Fig:crossvarVW} shows the variation with latitude of the variance
$(\langle T_1T_2\rangle-\langle T_1\rangle\langle T_2\rangle)$ and with
$T_1$, $T_2$, the antenna temperature at bands V and W of WMAP (61 and 94 GHz)
(Bennett et al. 2003a). 
This variance is free of instrumental noise because it corresponds to the 
cross-correlation of two independent maps. 
Each point in the plot represents the cross-variance
in the ring for all galactic longitudes 
and $\Delta (\sin b)=0.01$ (which comprises around 
15000 pixels). We used the data 
with no foreground subtraction with the kp2 mask (Bennett et al. 2003a),
which removes 15\% of the sky, mainly from the Galactic plane, 
the same mask that was used in the derivation
of their WMAP power spectrum. 
The figure  shows
two features: i) there is a significant decrease of
fluctuations toward the polar caps, especially for the southern cap; 
ii) there is a significant excess
 in the southern fluctuations ($b<0$) 
with respect to the north fluctuations (this fact was already present in
COBE-DMR data; L\'opez-Corredoira 1999, subsecc. 3.4).
The least squares linear fits ($y=a+c|b|$) to the ranges 
$b<-20^\circ $ and $b>20^\circ $ are respectively:
 
\begin{equation}
\sigma ^2_{cross,b<-20^\circ }(|b|)=(8.99\pm 0.28)\times 10^{-3}
\end{equation}\[
-(27.8\pm 5.9)\times 10^{-6}|b|(deg.) \ {\rm mK^2}
\]\[
\sigma ^2_{cross,b>20^\circ }(|b|)=(7.74\pm 0.24)\times 10^{-3}
\]\[
-(16.2\pm 5.1)\times 10^{-6}|b|(deg.) \ {\rm mK^2}
.\]

That is, $c=0$ is excluded within a 4.7-$\sigma $ level for the south
and 3.2-$\sigma $ for the north for the first feature, 
and $\sigma ^2_{cross}(b=-20^\circ )-\sigma ^2_{cross}(b=20^\circ )=
(1.01\pm 0.40)\times 10^{-3}$ mK$^2$, i.e. the second feature.
Equality between north and south is excluded to within 2.5-$\sigma $.

Bennett et al. (2003b, Fig. 10) point out that in the W bands
the Galactic contribution constitutes around 10\% of the power spectrum.
If we examine the same
dependence in the same maps once the foregrounds have been cleaned according
to their estimations, the results are (see also Fig. \ref{Fig:crossvarVW}):
\begin{equation}
\sigma ^2_{cross,b<-20^\circ }(|b|)=(8.59\pm 0.27)\times 10^{-3}
\end{equation}\[-(20.6\pm 5.6)\times 10^{-6}|b|(deg.) \ {\rm mK^2}
\]\[
\sigma ^2_{cross,b>20^\circ }(|b|)=(6.88\pm 0.23)\times 10^{-3}
\]\[+(2.2\pm 4.7)\times 10^{-6}|b|(deg.) \ {\rm mK^2}
,\]
and $\sigma ^2_{cross}(b=-20^\circ )-\sigma ^2_{cross}(b=20^\circ )=
(1.25\pm 0.38)\times 10^{-3}$ mK$^2$. The latitude dependence on the south
is less significant ($c=0$ excluded at 3.7-$\sigma $) and disappears 
in the north, but the south-north difference  is higher than before 
(3.3-$\sigma $).

These higher fluctuations in the southern galactic hemisphere have
also been observed by Eriksen et al. (2004a), Hansen et al. (2004a,b),
Chyzy et al. (2005) or Bernui et al. (2006) 
with higher asymmetry for lower latitudes (Chyzy et al. 2005).
The higher asymmetry for lower values of $|b|$ is expected from
the Galactic contamination because the dust emission is roughly
proportional to $1/\sin (|b|)$.

\subsection{Simulations}

We now do the same measures in 20 random realizations of the expected
WMAP sky (Eriksen et al. 2004b, 2005) (i.e., 20 measures of the 
excess south/north, and 40 measures of the slope).

The average slope $c$ (of a linear law $\sigma ^2(|b|)=a-c|b|$)
is $+11\times 10^{-6}$ $mK^2/deg.$ and the r.m.s. is $21\times 10^{-6}$
$mK^2/deg$, so our measured values of $+27.8 \times 10^{-6}$ $mK^2/deg$
and $+16.2\times 10^{-6}$ $mK^2/deg$ for the south and the north respectively 
are indeed within the expected values only for cosmological fluctuations.
The fact that we have a non-null value for the average $c$ ($<c>=+11.3\pm 
3.3\times 10^{-6}$) is due to the way the fluctuations are measured in
different rings with different latitudes; all rings have the same number
of points but different shapes; for instance, toward the polar caps
the areas are different from the rings near the plane.

The difference south/north in the same realizations
is on average $0.33\pm 0.28 \times 10^{-3}$ 
mK$^2$ (compatible with zero, which is the exact expected value, since
there are no reasons for an asymmetry north/south) and the r.m.s.
is $1.3 \times 10^{-3}$ mK$^2$. Again, our measured value in WMAP
of $1.25\times 10^{-3}$ mK$^2$ is within the expected statistical values.

\subsection{Latitude dependence on small angular scales}
\label{.sangsca}

We might think that the variations among zones could be reduced
if we used only a high-$l$ (small angular scale) 
range of the power spectrum with the
same shape for all selected regions instead of circular rings of variable
thickness and height parallel to the Galactic equator.
If we do the analysis of $[l(l+1)C_l/(2\pi )]$ for $316\le l \le 354$
with 28 points distributed with different
galactic longitudes and latitudes $\pm 36^\circ ,\pm 54^\circ,
\pm 72^\circ, \pm 90^\circ $ in circles of radius of 9$^\circ $ which
are not overlapping to each other, we get 

\begin{equation}
[l(l+1)C_l/(2\pi )](|b|=90^\circ )=3.0(\pm 0.6)\times 10^{-3} {\rm
mK^2/deg.} 
\label{pk90ss}
\end{equation}
\begin{equation}
\frac{d[(l(l+1)C_l/(2\pi )]}{d|b|}=+3(\pm 10) \times 10^{-6} {\rm
mK^2/deg.} 
,\end{equation}
that is, the error bar is larger than the measured slope.

\subsection{Constraints on the maximum Galactic contamination}
\label{.constrang}

{\bf All angles:}
We are therefore very limited in ascertaining the Galactic contamination
through this kind of analysis. The values of the present simulations allow us
to put a higher limit but these numbers permit a very wide range of
Galactic contamination amplitudes.
On average, for both hemispheres, it should be less than
$+43 \times 10^{-6}$ $mK^2/deg.$ (1$\sigma $) [$+64 \times 10^{-6}$ 
$mK^2/deg.$(2$\sigma $)].
Since Bennett et al.'s (2003b) foreground corrections 
reduce the value of the slope
$c$ by an average of 13$\times 10^{-6}$ $mK^2/deg$ and they claim that
this is representative of 10\% of contamination by the Galaxy,
an estimate of the maximum limit of contamination by the Galaxy
is less than 33\% (1$\sigma $) [49\% (2$\sigma $)]. 

Another way to analyse it would be to
suppose that the Galactic contamination follows
a cosecant law in the dependence on galactic latitude.
Then $c_{gal}/a_{gal}=9.5\times 10^{-3}$ (derived from a linear fit
in the same conditions as for the analysis of the WMAP data);
so the excess over the average in the simulation
($<c>=+11.3\times 10^{-6}$) in the
south of $c_{gal}=\Delta c=27.8\times 10^{-6}-11.3\times 
10^{-6}\pm 21\times 10^{-6}\
mK^2/deg.=16.5\pm 21 \times 10^{-6}\ mK^2/deg.$ implies that
$a_{gal, south}=1.7\pm 2.2\times 10^{-3}$, i.e. $19\pm 25$\% of the
total emission ($a=8.9\times 10^{-6}$). This means that the
Galactic emission in the south is less than 44\% (1$\sigma $) 
[69\% (2$\sigma $)]. 
The same analysis for the northern hemisphere gives:
less than 35\% (1$\sigma $) [64\% (2$\sigma $)]. Given that,
as said above with regard to the asymmetry south/north, 
the contamination in the south should be 1.94 times the contamination
in the north, we can further reduce the constraint in the north:
less than 26\% (1$\sigma $) [41\% (2$\sigma $)]. Therefore, on average
in the north and the south, the contamination is less than 35\% 
(1$\sigma $) [55\% (2$\sigma $)].

{\bf Small angular scales:}
With the numbers at small angular scales given in \S \ref{.sangsca},
we can set the constraint:
$[(l(l+1)C_l/(2\pi )](|b|=20^\circ )-[l(l+1)C_l/(2\pi )](|b|=90^\circ )]<0.49 \times 10^{-3}$ mK$^2$/deg
(1$\sigma $) [$<1.2\times 10^{-3}$ mK$^2$/deg (2$\sigma $)], which implies, assuming a
dust emission proportional to $1/sin(|b|)$, that $[l(l+1)C_l/(2\pi )](|b|=90^\circ )<2.5\times 10^{-4}$ (1$\sigma $)
[$<6.2\times 10^{-4}$ (2$\sigma $)].
Comparing with eq. (\ref{pk90ss}), we get the constraint
that the Galactic emission must be lower than 8\% (1$\sigma $)
[20\% (2$\sigma $)].
We see, then, that the restriction of values of $l$ is more
efficient for constraining the Galactic contamination.

\section{The frequency dependence}
\label{.freqdep}

In my opinion, the most important proof presented to claim that
most of off-plane microwave anisotropies have a cosmological origin
is the frequency dependence. The near independence in
the range of 50--90 GHz is not conclusive because this could
be produced by the Galaxy too, mainly as a combination of thermal dust and
rotational dust emission (L\'opez-Corredoira 1999). 
However, the surveys with a larger
range of frequencies, such as that at 170 GHz by Ganga et al. (1993),
BOOMERANG (Netterfield et al. 2002)
or ARCHEOPS (Tristam et al. 2005), have shown that 
the power spectrum is practically the same in amplitude and shape with
respect to WMAP measures and is constant over the frequency 
range 50--250 GHz. 
The spectral analysis of FIRAS+WMAP by Fixsen (2003) also shows
that the anisotropies are Planckian, although Fixsen (2003) assumes
that the frequency and spatial dependences of the Galaxy are separable
and modelled by templates, which is not a trustworthy assumption.
Since Galactic dust thermal emission of 7--20 K temperature grains
should have some frequency dependence, we must conclude that
they are not predominant. We have, within the error ranges of
the calibration and within the error bars of the power spectrum
(around a 10\% in total), a constant amplitude over a factor 5 in frequency.

The amplitude of the anisotropies depends on two factors (L\'opez-Corredoira
1999): the variation of the mean flux with frequency, and the ratio
of diffuse and cloud emission as a function of frequency; but in any
case, the variation of $(\Delta T)^2$ should be at least twice as high at
250 GHz than at 50 GHz. This minimum factor roughly equal to two stems 
from the multiplication of the first two factors specified 
in L\'opez-Corredoira (1999, eq. 2):
\begin{equation}
\frac{(\Delta T)^2(\nu =250\ {\rm GHz})}{(\Delta T)^2(\nu =50\ {\rm GHz})}=
\left \langle \frac{T (\nu =250\ {\rm GHz})}{T (\nu =50\ {\rm GHz})}\right \rangle ^2
\end{equation}\[
\times \left \langle \frac{\frac{\delta T}{T}(\nu =250\ {\rm GHz})}
{\frac{\delta T}{T}(\nu =50\ {\rm GHz})}\right \rangle ^2
.\]
$\left \langle \frac{T (\nu =250\ {\rm GHz})}{T (\nu =50\ {\rm GHz})}
\right \rangle \approx 5$ (L\'opez-Corredoira 1999, Fig. 5), and 
$\left \langle \frac{\frac{\delta T}{T}(\nu =50\ {\rm GHz})}
{\frac{\delta T}{T}(\nu =250\ {\rm GHz})}\right \rangle <\approx 3.3$ 
(L\'opez-Corredoira 1999, eq. 12) so 
$\frac{(\Delta T)^2(\nu =250\ {\rm GHz})}
{(\Delta T)^2(\nu =50\ {\rm GHz})}>\approx 2.3$.

Assuming this roughly minimum factor of two in the increase in
Galactic anisotropy at 250 GHz with respect to 50 GHz, if we observe a
maximum of 10\% excess in the total contribution, it means that the
Galactic emission should be lower than a 10\% of the total contribution 
of the anisotropies in average in the range of frequencies.

\section{Other contaminants}
\label{.others}

Although it is not the matter for this paper, it is important to
point out that, apart from the cosmological and the Galactic signal,
there may other contaminants: either from closer sources (in the
solar system or the solar neighbourhood), 
or extragalactic sources much closer than $z=1000$--1500
(as  is supposed for the cosmological origin).

Copi et al. (2006, 2007) show that the two lowest cosmologically interesting 
multipoles, $l$=2 and 3, are not statistically isotropic. 
The planes of the quadrupole and the octopole are unexpectedly aligned
(de Oliveira-Costa 2004; Copi et al. 2006, 2007; Land \& Magueijo 2007;
Raki\'c \& Schwarz 2007). 
Indeed, the combined quadrupole plus octopole is surprisingly 
aligned with the geometry and direction of motion of the solar system: 
the plane they define is perpendicular to the ecliptic plane and to 
the plane defined by the dipole direction, and the ecliptic plane 
carefully separates stronger from weaker extrema, running within a 
couple of degrees of the null-contour between a maximum and a minimum over 
more than 120$^\circ $ of the sky. The axis of maximum asymmetry of the 
WMAP data tends to lie close to the ecliptic axis (Eriksen et al. 2004a).
There is alignment of the quadrupole and octopole with each other and 
they are correlated with each other: 99.6\% C.L.(Copi et al. 2007). 
There are also statistically significant correlations with local geometry, 
especially that of the solar system ($>99.9$\% C.L.; Copi et al. 2006). 
Moreover, the angular two-point correlation 
function at scales $>60$ degrees in the regions outside the Galactic cut 
is approximately zero in all wavebands 
and is discrepant with the best fit $\Lambda $CDM inflationary model: 
99.97\% C.L. for the discrepancy (Copi et al. 2007; Raki\'c \& Schwarz 2007). 
Eriksen et al. (2004a) also found that the ratio of the 
large-scale fluctuation amplitudes in the southern ecliptic hemisphere is 
high at the level 98--99\% , with an absence of large-scale power in the vicinity
of the north ecliptic pole. This asymmetry is stable with respect
to frequency and sky coverage. This seems to point to the presence of some contamination
in relation with the solar system or neighbourhood,
although part of the effect may be due to non-uniform observational time 
sky coverage (Chyzy et al. 2005). 
Abramo et al. (2006) claim that if a hypothetic foreground 
produced by a cold spot in the Local Supercluster is subtracted 
from the CMB data, the amplitude of the 
quadrupole is substantially increased, and the statistically 
improbable alignment of the quadrupole with the octopole is 
substantially weakened, but this does not explain the coincidence
of the alignment with the ecliptic.

The extragalactic contamination is still not very clear. In this respect,
the discovery that the hard X-ray background
measured by the HEAO-1 satellite and the number counts of radio galaxies in the 
NVSS survey are correlated with the WMAP microwave fluxes
(Boughn \& Crittenden 2004) should be pointed out. One could raise the objection that
microwave anisotropies are produced over a very wide range of angles,
which include structures of several degrees and largest structures
while galaxies and clusters are generally smaller. 
However, if one considers the
very large structures of filaments, walls, hyperclusters, etc. one
can find in the sky structures of even several tens of degrees; that
is, excess of densities of galaxies over regions with several tens of degrees.
As a matter of fact, Narlikar et al. (2003) explain the
peak at $l=200$ and other peaks too in the power spectrum 
in terms of rich clusters of galaxies.
The galaxy counts in L\'opez-Corredoira \& Betancort-Rijo
(2004, Fig. 1) also suggest structures of this kind and size. I did the
cross correlation of these near-infrared galaxy counts with COBE-DMR-90 GHz
and  found nothing significant (cross-corr. for $|b|>20^\circ $:
$5.5(\pm 11.2)\times 10^{-7}$ K).
However, as said in \S \ref{.gauss}, using the WMAP data and 2MASS galaxy 
catalog, Cao et al. (2006) show that the 
non-gaussianity of the 2MASS galaxies is imprinted on WMAP maps,
and Cabre et al. (2006) cross-correlate the third-year-WMAP data 
with optical galaxy samples extracted from the SDSS-DR4 
giving a positive signal-to-noise of about 4.7.

Another consideration that could point to the importance of the 
non-cosmological extragalactic contamination over photons of cosmological
origin is that the simulations of gravitational lensing of the microwave background 
by galaxy clusters at $z<1$ under any plausible Big Bang model variation 
produces far more dispersion in the angular size of the primary acoustic peaks 
than WMAP observations allow (Lieu \& Mittaz 2005). 
When all the effects are taken together, it is difficult to understand how WMAP could 
reveal no evidence whatsoever of lensing by groups and clusters. 
Cool spots in the microwave background are too uniform in size to have 
travelled from $z=1000$--1500 to us. 
There should be a spread of sizes around the average, 
with some of these cool spots noticeably larger and others noticeably smaller. 
But this dispersion of sizes is not seen in the data. 
Too many cool spots have the same size. Moreover,
the observed WMAP Sunyaev--Zel'dovich effect caused by the clusters only 
accounts for about 1/4 of the expected decrement (Lieu et al. 2006);
although the level of Sunyaev--Zel'dovich effect is observed
as in the predictions in radio (Bonamente et al. 2006).
Effects like a central cooling flow in clusters, 
the abundance of hot cluster gas, large scale radial decline 
in the temperature, uncertainties in the $\beta$-model
and the role played by cluster radio sources are too weak to change the 
estimation (Lieu et al. 2006).
Under Big Bang premises, 
this implies that the cosmological parameters (including the Hubble constant, 
the amount of dark matter, etc.) used to predict the original, pre-lensed 
sizes of the cool and hot spots in the microwave background might be 
wrong or some of these cool spot structures are caused by nearby physical 
processes and are not really remnants of the creation of the Universe; 
or there is some other, unknown factor damping the effects of dispersion 
and focusing. It was speculated that the large-scale curvature of space may 
not entirely be an initial value problem related to inflation. The absence of 
gravitational lensing of the CMB points to the possibility that even effects 
on light caused by wrinkles in the space of the late (nearby) Universe have 
been compensated for, beyond some distance scale, by a mechanism that 
maintains a flat geometry over such scales.
Or high energy electrons may 
synchrotron radiate in the intracluster magnetic field of strength 
B $<\sim$ 1$\mu$G to produce cluster microwave emissions in the 
WMAP passbands that account for the missing Sunyaev--Zel'dovich effect flux
(Lieu \& Quenby 2006).
However, one is also tempted to interpret this in terms of the importance
of extragalactic non-cosmological contribution in the microwave anisotropies.

\section{Summary and conclusions}

Summing up, CMBR anisotropies are not totally free from
from solar system, Galactic or extragalactic contamination, and an accurate
way of correcting for all these contributions has still to be devised since we do not
have accurate information on the microwave emission of any of these
contributions. 

Considering only the Galactic dust component, the main
topic of this paper, it was shown that the methods used to remove 
it (templates, cross-correlations, assumption of a Galactic power
spectrum, MEM, ILC, etc.) are all  inaccurate and one should
not expect to produce maps clean of Galactic dust contamination by applying
them. The analysis of galactic latitude dependence of these anisotropies
and the fact that the power spectrum
is nearly independent of the frequency over a range 50--250 GHz
can be considered at least as a proof that the Galactic dust emission
is lower than a 10\% on the $\sim 1$ degree scale 
[or double of this value considered to 2$\sigma $], possibly higher
for lower $l$ multipoles.
This uncertainty in the Galactic contamination 
may produce important systematic errors in some cosmological
parameters.  
The determination of the error bars in the cosmological
parameters taking into account the possible foreground contamination
values is beyond the scope of this paper. Further research is needed in this respect.
In any case, one thing is clear: the present error bars calculated
for the cosmological parameters (e.g., Spergel et al. 2007)
are very significantly underestimated
and the range of possible values is not as small as indicated by
the ``precision cosmology'' claim.

Therefore, I conclude that the most pessimistic possibilities among those
claimed in L\'opez-Corredoira (1999) (that Galactic contamination could be 
dominant) is discarded, but I keep my position that we are far from achieving
a ``precision cosmology'' with CMBR data.
My suggestion with regard to the problem would be that rather than performing an
incorrect subtraction of the foreground and claiming that a clean map
is obtained that is purely cosmological, 
we might better use the map without corrections or only with a
first-order approximation subtraction, and calculate the cosmological
parameters with the appropriate error bars taking into account the
uncertainties in the foregrounds (Galactic or any other), 
to be constrained only by the frequency dependence.

\vspace{10mm}

{\bf Acknowledgments:} 
Thanks are given to J. A. Rubi\~no-Mart\'\i n 
(IAC, Tenerife, Spain)
for providing some numerical codes to read HEALPix format data and
the calculation of the power spectrum; to H. K. Eriksen
(Inst. Theor. Astroph., Oslo, Norway) for providing me the
maps of the simulations derived in his paper Eriksen et al. (2005);
to C. M. Guti\'errez (IAC, Tenerife, Spain) and the anonymous referee
for helpful comments on this paper; and to T. J. Mahoney (IAC, Tenerife,
Spain) for proof-reading of this paper.


\begin{thebibliography}{99}

\bibitem{} Abramo, L. R., Sodre, L. Jr., \& Wuensche, C. A. 2006,
Phys. Rev. D, 74(8), 3515
 
\bibitem{} Bennett, C. L., Halpern, M., Hinshaw, G., et al. 2003a, ApJS,
148, 1 

\bibitem{} Bennett, C. L., Hill, R. S., Hinshaw, G., et al. 2003b,
ApJS, 148, 97

\bibitem{} Bernard, J. P., Abergel, A., Ristorcelli, I., et al. 1999, A\&A,
347, 640

\bibitem{} Bernui, A., Mota, B., Reboucas, M. J., \& Tavakol, R. 2007,
A\&A 464, 479

\bibitem{} Bernui, A., Tsallis, C., \& Villela, T. 2007,
astro-ph/0703708

\bibitem{} Bernui, A., Villela, T., Wuensche, C. A., Leonardi, R.,
\& Ferreira, I. 2006, A\&A, 454, 409

\bibitem{} Bonamente, M., Joy, M. K., LaRoque, S. J., Carlstrom, J. E., 
Reese, E. D., \& Dawson, Kyle S. 2006, ApJ 647, 25
 
\bibitem{} Boughn, S., \& Crittenden, R. 2004, Nature, 427, 45

\bibitem{} Cabr\'e, A., Gazta\~naga, E., Manera, M., Fosalba, P., 
\& Castander, F. 2006, MNRAS, 372, L23

\bibitem{} Cao, L., Chu, Y.-Q., \& Fang, L.-Z. 2006, MNRAS, 369, 645

\bibitem{} Casassus, S., Readhead, A. C. S., Pearson, T. J., Nyman, L.-A., 
Shepherd, M. C., \& Bronfman, L. 2004, ApJ, 603, 599

\bibitem{} Chiang, L.-Y., Naselsky, P. D., \& Coles, P., 2007,
ApJ, 664, 8

\bibitem{} Chiang, L.-Y., Naselsky, P. D., Verkhodanov, O. V.,
\& Way, M. J. 2003, ApJ, 590, 65

\bibitem{} Chyzy, K. T., Novosyadlyj, B., \& Ostrowski, M. 2005,
astro-ph/0512020

\bibitem{} Coles, P., Dineen, P., Earl, J., \& Wright, D. 2004,
MNRAS, 350, 983

\bibitem{} Copi, C. J., Huterer, D., Schwarz, D. J., \& Starkman, G. D.
2006, MNRAS, 367, 79

\bibitem{} Copi, C. J., Huterer, D., Schwarz, D. J., \& Starkman, G. D.
2007, Phys. Rev. D, 75(2), 3507

\bibitem{} Coulson D., Ferreira P., Graham P., \& Turok N. 1994, Nature,
368, 27

\bibitem{} Davies, R. D., Guti\'errez, C. M., Hopkins, J., et al. 1996,
MNRAS, 278, 883

\bibitem{} Draine B. T., \& Lazarian A. 1998, ApJ, 494, L19

\bibitem{} Draine B. T., \& Lazarian A. 1999, ApJ, 512, 740

\bibitem{} Dupac, X., Bernard, J. P., Boudet, N., Giard, M.,
Lamarre, J. M., M\'eny, C., Pajot, F., \& Ristorcelli, I. 2004, in:
Multiwavelength Cosmology (Astroph. Space Sci. library, vol. 301),
Kluwer Academic Publishers, Dordrecht (The Netherlands), p.89

\bibitem{} Eriksen, H. K., Hansen, F. K., Banday, A. J., G\'orski, K. M.,
\& Lilje, P. B. 2004a, ApJ, 605, 14

\bibitem{} Eriksen, H. K., Banday, A. J., G\'orski, K. M., \& Lilje, P. B.
2004b, ApJ, 612, 633

\bibitem{} Eriksen, H. K., Banday, A. J., G\'orski, K. M., \& Lilje, P. B.
2005, astro-ph/0508196

\bibitem{} Fern\'andez-Cerezo, S., Guti\'errez, C. M., Rebolo, R., et al.
2006, MNRAS, 370, 15

\bibitem{} Ferreira P. G., G\'orski K. M., 
\& Magueijo J. 1998, ApJ, 503, L1

\bibitem{} Finkbeiner, D. P., Davis, M., \& Schlegel, D. J. 1999,
ApJ, 524, 867

\bibitem{} Finkbeiner, D. P., Langston, G. I., \& Minter, A. H. 2004,
ApJ, 617, 350

\bibitem{} Fixsen, D. J. 2003, ApJ, 594, L67

\bibitem{} Ganga, K., Cheng, E., Meyer, S., \& Page, L. 1993,
ApJ, 410, L57

\bibitem{} Gutierrez de La Cruz, C. M., Davies, R. D., Rebolo, R., 
Watson, R. A., Hancock, S., \& Lasenby, A. N. 1995, MNRAS, 442, 10

\bibitem{} Guti\'errez, C. M., Rebolo, R., Watson, R. A., Davies, R. D., 
Jones, A. W., \& Lasenby, A. N. 2000, ApJ, 529, 47

\bibitem{} Halverson, N. W., Leitch, E. M., Pryke, C., et al. 2002,
ApJ, 568, 38

\bibitem{} Hammersley, P. L., Garz\'on, F., Mahoney, T., \& Calbet, X.
1995, MNRAS, 273, 206

\bibitem{} Hansen, F. K., Banday, A. J., \& G\'orski, K. M. 2004a, MNRAS,
354, 641

\bibitem{} Hansen, F. K., Balbi, A., 
Banday, A. J., \& G\'orski, K. M. 2004b, MNRAS, 354, 905

\bibitem{} Hansen, F. K., Banday, A. J., Eriksen, H. K., G\'orski, K. M.,
\& Lilje, P. B. 2006, ApJ 648, 784

\bibitem{} Herbstmeier, U., \'Abrah\'am, P., Lemke, D., et al. 1998,
A\&A, 332, 739

\bibitem{} Kiss, C., \'Abrah\'am, P., Klaas, U., Lemke, D., 
H\'eraudeau, P., del Burgo, C., \& Herbstmeier, U. 2003, A\&A, 399, 177

\bibitem{} Kogut A., 
Banday A. J., Bennett C. L., et al. 1996a, ApJ, 460, 1

\bibitem{} Land, K., Magueijo, J. 2007, MNRAS, 378, 153

\bibitem{} Lawrence A. 2001, MNRAS, 323, 147

\bibitem{} Leitch, E. M., Readhead, A. C. S., 
Pearson, T. J., \& Myers, S. T. 1997, ApJ, 486, L23 

\bibitem{} Leitch, E. M., Readhead, A. C. S., Pearson, T. J., Myers, S. T.,
Gulkis, S., \& Lawrence, C. R. 2000, ApJ, 532, 37

\bibitem{} Lieu, R., \& Mittaz, J. P. D. 2005, ApJ, 628, 583

\bibitem{} Lieu, R., Mittaz, J. P. D., \& Zhang, S.-N. 2006, 
ApJ, 648, 176

\bibitem{} Lieu, R., \& Quenby, J. 2006, astro-ph/0607304

\bibitem{} Liu, X., \& Zhang, S. N. 2005, ApJ, 633, 542

\bibitem{} Liu, X., Zhang, S. N. 2006, ApJ, 636, L1

\bibitem{} L\'opez-Corredoira, M. 1999, 
A\&A, 346, 369

\bibitem{} L\'opez-Corredoira, M., Cabrera-Lavers, A., Garz\'on, F.,
\& Hammersley, P. L. 2002, A\&A, 394, 883

\bibitem{} L\'opez-Corredoira, M., \& Betancort-Rijo, J., 2004,
A\&A, 416, 1

\bibitem{} Magueijo, J. 2000a, ApJ, 528, L57

\bibitem{} Magueijo, J. 2000b, ApJ, 532, L157

\bibitem{} Masi, S., Ade, P. A. R., Bock, J. J., et al. 2001,
ApJ, 553, L93

\bibitem{} McEwen, J. D., Hobson, M. P., Lasenby, A. N., \& Mortlock, D. J.
2006, MNRAS, 371, L50

\bibitem{} Mukherjee, P., Jones, A. W., Kneissl, R., \& Lasenby, A. N. 2001,
MNRAS, 320, 224	

\bibitem{} Narlikar, J. V., Vishwakarma, R. G., Hajian, A., 
Souradeep, T., Burbidge, G., \& Hoyle, F. 2003, ApJ, 585, 1

\bibitem{} Naselsky, P. D., Novikov, I. G., \& Chiang, L.-Y. 2006,
ApJ, 642, 617

\bibitem{} Netterfield, C. B., Ade, P. A. R., Bock, J. J., et al.
2002, ApJ, 571, 604

\bibitem{} Oliveira-Costa, A. de, Kogut, A., 
Devlin, M. J., et al. 1997, ApJ, 482, L17

\bibitem{} Oliveira-Costa, A. de, Tegmark, M., 
Page, L. A., \& Boughn, S. P., 1998, ApJ, 509, L9

\bibitem{} Oliveira-Costa A. de, Tegmark, M.,
Guti\'errez, C. M., Jones, A. W., Davies, R. D.,
Lasenby, A. N., Rebolo, R., \& Watson, R. A. 1999,
ApJ, 527, L9

\bibitem{} Oliveira-Costa, A. de, Tegmark, M., Finkbeiner, D. P., et al.
2002, ApJ, 567, 363

\bibitem{} Oliveira-Costa, A. de, Tegmark, M., Zaldarriaga, M.,
\& Hamilton, A. 2004, Phys. Rev. D, 69(6), 3516

\bibitem{} Oliveira-Costa, A. de, \& Tegmark, M. 2006, Phys. Rev. D, 74(2),
3005

\bibitem{} Paladini, R., Montier, L., Giard, M., Bernard, J. P., Dame, T. M.,
Ito, S., \& Mac\'\i as-P\'erez, J. F. 2007, A\&A, 465, 839

\bibitem{} Pando, J., Valls-Gabaud, D., 
\& Fang, L.-Z., 1998, Phys. Rev. Lett., 81(21), 4568

\bibitem{} Park, C.-G. 2004, MNRAS, 349, 313

\bibitem{} Raeth, C., Schuecker, P., \& Banday, A. J. 2007, MNRAS 380, 466

\bibitem{} Raki\'c, A., \& Schwarz, D. J. 2007, Phys. Rev. D, 75(10), 3002

\bibitem{} Robitaille, P.-M., 2007, Progress in Physics 1/2007, 3

\bibitem{} Rubi\~no-Mart\'\i n, J. A., Aliaga, A. M., Barreiro, R. B.,
et al. 2006, MNRAS, 369, 909 

\bibitem{} Schlegel, D. J., Finkbeiner, D. P., \& Davis, M. 1998,
ApJ, 500, 525

\bibitem{} Spergel, D. N., Bean, R., Dore, O., et al. 2007,
ApJS 170, 377
 
\bibitem{} Tegmark M., 1998, ApJ 502, 1

\bibitem{} Then, H., 2006, MNRAS, 373, 139

\bibitem{} Tojeiro, R., Castro, P. G., Heavens, A. F., \& Gupta, S.
2006, MNRAS, 365, 265

\bibitem{} Tristam, M., Patanchon, G., Mac\'\i as-P\'erez, J. F.,
et al. 2005, A\&A, 436, 785

\bibitem{} Unavane M., Gilmore G., Epchtein N., Simon G.,
Tiph\`ene D., Batz B. de, 1998, MNRAS, 295, 119

\bibitem{} Verschuur, G. L., 2007, arXiv:0704.1125

\bibitem{} Vielva, P., Mart\'\i nez-Gonz\'alez, E., Barreiro, R. B.,
Sanz, J. L., \& Cay\'on, L. 2004, ApJ, 609, 22

\bibitem{} Watson, R. A., Rebolo, R., Rubi\~no-Mart\'\i n, J. A.,
Hildebrandt, S., Guti\'errez, C. M., Fern\'andez-Cerezo, S.,
Hoyland, R. J., \& Battistelli, E. S. 2005, ApJ, 624, L89

\end{thebibliography}
\end{document}